# Siberian wildfires and related gas and aerosol emissions into the atmosphere over the past decades

I.I. Mokhov[1,2], S.A. Sitnov[1]

[1]A.M. Obukhov Institute of Atmospheric Physics, Russian Academy of Sciences
[2]Lomonosov Moscow State University
mokhov@ifaran.ru

**Abstract**

Siberian wildfires and related regional anomalies of atmospheric impurities during the period of 2000-2019 are analyzed. The long-range transport of biomass burning products from Siberian wildfires into the Arctic atmosphere during the period of 2000-2019 is estimated. An analysis of the characteristics of forest fires over the past two decades revealed a significant increase in radiation power of an average Siberian wildfire. A joint analysis of fire activity in Siberian forests, as well as the contents of the black carbon (BC) and carbon monoxide (CO) contents in the Arctic atmosphere, indicates that extreme fire events cause the development of regional anomalies in BC and CO. Correlation between the anomalies of BC (CO) over the Russian segment of the Arctic and the number of Siberian wildfires is found to be statistically significant. Using a linear regression, an estimates of the sensitivity of the total BC content and in the volume mixing ratio of CO to the increase in the number of fires. The results of an analysis of the long-range BC transport into the Arctic It is shown, in particular, that the transport of BC to the Arctic from the Siberian regions with fires in the summer of 2019 was associated with the features of large-scale atmospheric circulation characteristic for atmospheric blocking.

**Introduction**

Forests play an important role in the global carbon cycle and regulate the climate. Fire activity has a significant impact on biodiversity and ecosystem resilience. Natural fires are an important source of greenhouse gases and aerosol emissions into the atmosphere. An analysis of long-term observations of fire activity in boreal forests indicates an increase in the annual number of fires in recent decades and an earlier onset of the fire hazard period. According to model estimates, with continued global warming in the regions of Northern Eurasia, we should also expect an increase in natural fires and an increase in the duration of the fire hazard period [1-8].

The formation of natural fires under warming is facilitated by the tendency to reduce the precipitation during fire seasons in the mid-latitude regions, in particular, in Eurasian regions. In addition, weather and climate anomalies characterized by dry conditions of summer seasons in mid-latitudes and increased fire hazard are associated with blocking anticyclones (blockings) in the troposphere. The influence of blockings manifests itself in regional anomalies with significant differences in meteorological conditions in adjacent areas and significant interannual variability. It is significant that, according to model estimates, warming leads to an increase in the probability of longer atmospheric blockings [9-12].

Climate warming, which is most noticeable in high latitudes (Arctic amplification) [13,14], leads to a change in the boreal forest fire regimes with an increase in the risk of catastrophic fires, in particular in Siberia. Atmospheric transport of aerosol products of fires, including the black carbon (BC), which effectively absorbs solar radiation, to high latitudes contributes to additional heating of the Arctic system under global warming [15-18]. When deposited on snowy and icy surfaces, BC reduces the albedo and increases the absorption of solar radiation, increasing the heating of the underlying surface [19]. On the territory of Russia, the main sources of BC during the warm period of the year are massive forest fires, which account for more than 80% of the total BC emission [17]. Forest fires in western and eastern Siberia, respectively, account for about a quarter and about a half of the total number of forest fires in Northern Eurasia. In recent years, there has been an increase in the radiation power of an average Russian fire [10].

The transport of pollutants from middle latitudes to the Arctic is often associated with atmospheric blocking phenomena. In particular, seasonal variations in the aerosol concentration over Arctic Alaska are associated with seasonal variations in the frequency and position of mid-latitude blocking anticyclones [20]. Aerosols can be transported from south to north over the western periphery of quasi-stationary anticyclones.

This paper presents the results of an analysis of the features and changes in Siberian fires and related regional anomalies of atmospheric impurities over the past decades. Estimates of the relationship between BC and CO pollution in the Arctic atmosphere and Siberian forest fires are given with an analysis of possible mechanisms for the long-range transport of combustion products from Siberia to the Arctic.

**Data and methods used**

The characteristics of forest fires in Siberia were analyzed using MODIS satellite data (measurements on board the Terra and Aqua platforms) from November 2000 to December 2020 with a fire diagnostics accuracy of at least 80%. MODIS Active Fire Products (C6, L2) processed by the standard MCD14ML algorithm is available through the FIRMS (Fire Information for Resource Management System, https://earthdata.nasa.gov). A detailed description of the MODIS C6 fire detection algorithm is presented in [21].

Hourly data were used for the BC column mass density (M2T1NXAER v5.12.4) and BC mass concentration at the surface (M2T1NXAER v5.12.4), as well as data for the northerly wind at a level of 10 m (M2T1NXSL v.5.12.4 10Mrfedc) from the MERRA reanalysis -2.

MERRA-2 (Modern-Era Retrospective analysis for Research and Applications, version 2) is a NASA reanalysis with assimilation of various satellite measurements since 1980. MERRA-2 uses a version of GEOS-5 (Goddard Earth Observing System Model, version 5) data assimilation system for the synthesis of regular gridded time series data with a spatial resolution of 0.5° x 0.625° (latitude x longitude) at 72 pressure levels (from the surface to 0.01 hPa) both instantaneous and time-averaged (hourly, 3-hourly and monthly). The advantage of MERRA-2 over the previous version is associated with the joint assimilation of aerosol and meteorological fields, taking into account their interaction through the radiative effect of atmospheric aerosol [22].

The data for the CO mole fraction in the atmosphere (AIRS3STM_006_CO_VMR_A) from measurements using the AIRS (Atmospheric InfraRed Sounder, C6) high-resolution spectrometer aboard the Aqua satellite since September 2002. The accuracy of CO measurements is 15%, the spatial resolution in nadir is 13,5 km, vertical resolution is 1 km [23]. BC and CO data are available through the Giovanni Internet server (https://giovanni.gsfc.nasa.gov/giovanni) developed and maintained by the NASA Goddard Earth Sciences Data and Information Services Center [24].

Also, the data for geopotential height at the level of 500 hPa (H500) from the NCEP/NCAR reanalysis [25] were used. H500 daily data with a spatial resolution of 2.5° x 2.5° are available at (https://www.esrl.noaa.gov/psd/data/gridded/data.ncep.reanalysis.html).

The relationship between the content of BC and CO in the atmosphere over the Arctic and Siberian forest fires was analyzed. In particular, the results for the Arctic sector (66–90° N, 30–180° E) with an area of 9.2 million km$^2$ and for the Siberian regions (50–75°N, 60–140°E) with an area of 11.3 million km$^2$ are presented in this paper.

**Results and discussion**

*Siberian wildfires*

The results of the analysis of hotspots detected by MODIS data indicate a regular manifestation of the activity of Siberian forest fires with strong intra- and inter-annual variability. The regional number of wildfires in Siberian forests (N) and the total fire radiation power (FRP) are characterized by relatively weak positive trends. It should be noted that the estimates of the characteristics of N and FRP variations indicate the heterogeneity of the data and the impossibility to obtain statistically reliable estimates, including estimates of trends in the N and FRP changes.

On the other hand, the analysis of data for monthly variations in the radiation power of the average Siberian fire (FRPavg), estimated as the ratio of the total FRP to the total number of fires, indicates the possibility of obtaining statistically significant estimates (Fig. 1). Analysis of FRPavg variations in summer seasons for the period 2003-2019 shows that the radiation power of an average Siberian fire is characterized by a noticeable increase with a linear trend of 1.3 (±0.8) %/year (statistically significant at the level of 5%). The trend was estimated in % of the average FRPavg for the period 2003-2019. The obtained results confirm the tendency of increase in the radiation power of the Siberian forest fire previously estimated from shorter time series [10].

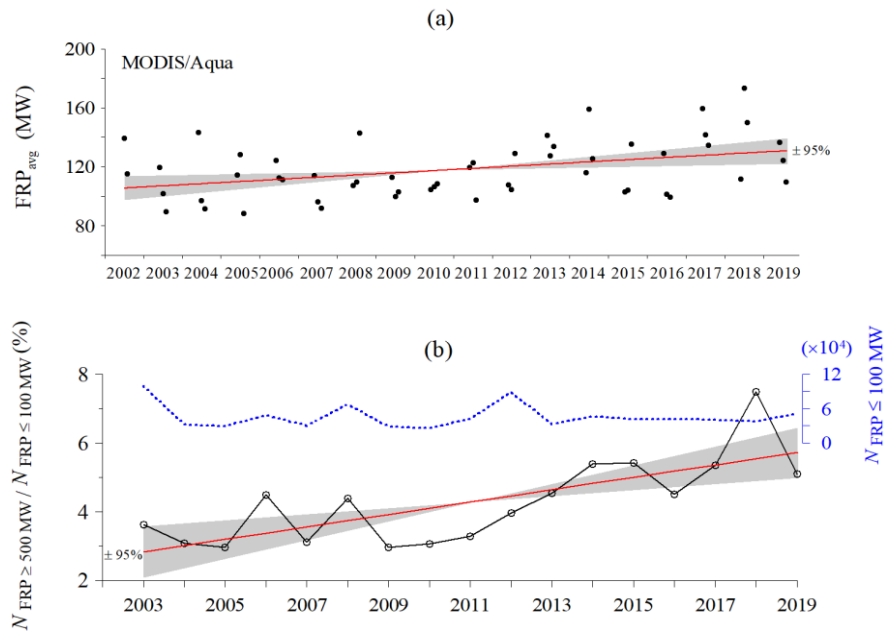

**Figure 1.** (a) Interannual variations of $FRP_{avg}$ in June-July-August during the period of 2002-2019 and its linear trend (red line) with a 95% confidence interval (shaded area). (b) The ratio of the number of intense fires (FRP ≥ 500 MW) to the number of weak fires (FRP ≤ 100 MW) (black line) and its linear trend (red line) with a 95% confidence interval (shaded area). Variations in the number of weak fires (blue dashed line, right scale) are also shown.

The total distribution of forest fires is dominated by hotspots characterized by a low FRP [10]. At the same time in [10] it was shown that from west to east in Russia the frequency of weak wildfires decreases while the frequency of intense forest fires increases. It is necessary to analyze the tendencies of long-term changes in Siberian wildfires of various intensities. Figure 1 shows the interannual variations in the ratio of the number of summer wildfires in Siberia with an intensity of more than 500 MW and less than 100 MW. The results obtained indicate that in the past two decades, in the territory of Siberia, there has been a predominant increase in the number of intense forest fires with a statistically significant linear trend of 1.8 ± 0.9% / 10 years. The long-term changes in the number of summer Siberian fires with a FRP less than 100 MW does not show any pronounced tendencies (Fig. 1).

*Relationship of the BC and CO anomalies in the Arctic atmosphere with Siberian wildfires*

Figure 2a shows the differences between the current monthly number of Siberian wildfires and corresponding long-term monthly mean values, which characterize the anomalies of fire activity in the forests of Siberia. The average monthly anomalies of BC and CO in the atmosphere over the Russian sector of the Arctic estimated in a similar way are shown in Figs. 2b and 2c, respectively. Comparison Fig. 2a and Fig. 2b indicates that the increased activity of forest fires in Siberia in August 2002, May 2003, April 2008, June–July 2012, and July–August 2019 (Fig. 2a) was accompanied by positive anomalies in the BC content in the Arctic atmosphere (Fig. 2b). Correlation of monthly BC anomalies in the Arctic with the number of Siberian fires in the period 2001-2019 reaches r = 0.70 (with a 95% confidence interval of 0.62-0.77) with an increase in the summer season up to r = 0.77 (with a corresponding confidence interval of 0.64-0.86). The high correlation between BC atmospheric content and forest fires is explained by the relationship between the emission of biomass combustion products into the atmosphere and the total area of forest fires [26]. The relationship between Arctic BC anomalies and Siberian forest fires is confirmed by the scatterplot (Fig. 2d). It can be seen from Fig. 2d that the increase in the BC content in the Arctic atmosphere is associated with an increase in fire activity in Siberian forests. Based on linear regression, an estimate of the sensitivity of the BC column density over the Russian sector of the Arctic to changes in the number of forest fires in Siberia is $1.9 \cdot 10^{-2}$ kg km$^{-2}$ / 1000 forest fires.

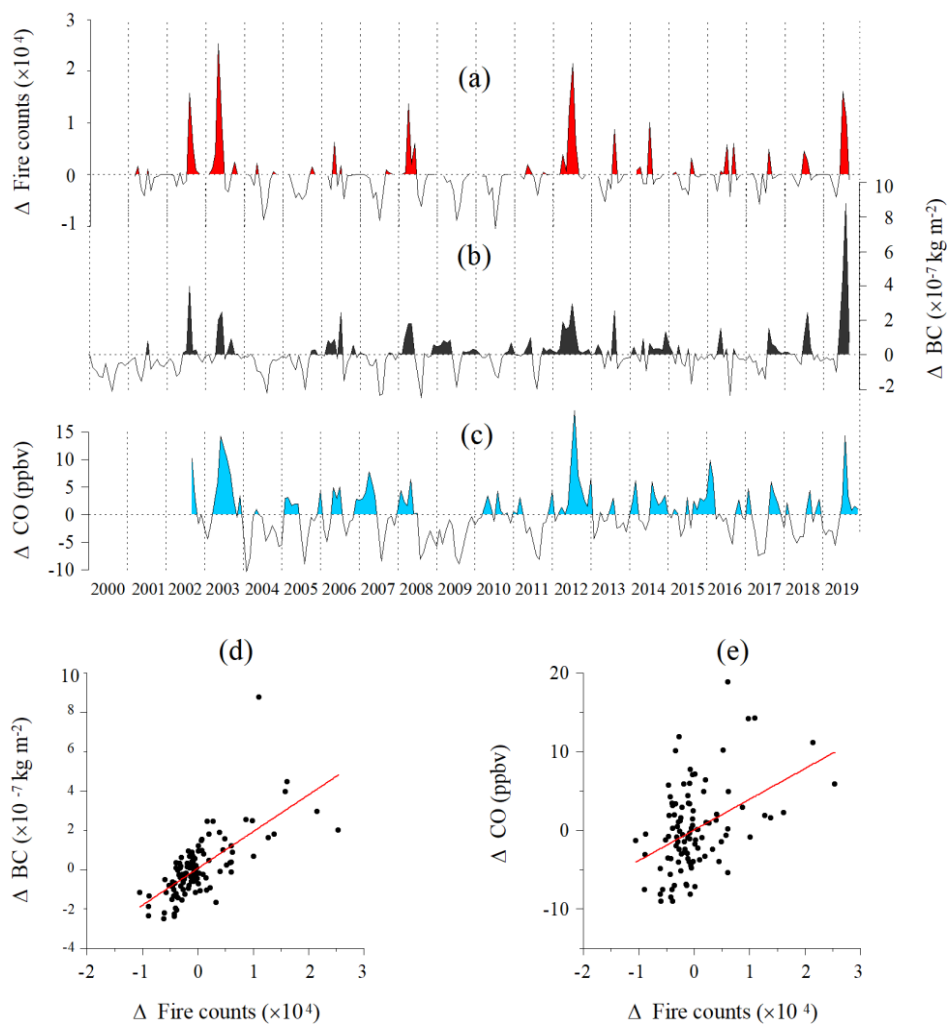

**Figure 2.** Monthly variations in the number of Siberian wildfires (a) and monthly variations in the BC column density (b) and in the CO volume mixing ratio at 700 hPa (c) in the Arctic atmosphere. Scatter plots between Siberian wildfires and BC column density in the Arctic (d) and between Siberian wildfires and CO volume mixing ratio at 700 hPa in the Arctic (e). The anomalies in (a-c) were obtained by subtraction of the long-term monthly means from the time-series. The scatter plots show relations between the anomalies in the fire hazardous periods (April to September).

Comparison of Fig. 2a and Fig. 2c shows that the increased fire activity of Siberian forests in 2003, 2012, and 2019 was also accompanied by positive CO anomalies in the Arctic troposphere (Fig. 2c). The correlation coefficient of the monthly mean values of CO anomalies in the Arctic with the number of fires in Siberia in the period 2002-2019 was obtained to be equal to 0.39 (with a 95% confidence interval of 0.26-0.51) with the increase in the summer season to $r = 0.48$ (0.24-0.67). The weaker correlation of CO with forest fires (in comparison with BC) can be associated with a greater variety of CO sources for the Russian sector of the Arctic. The results obtained indicate, in particular, the manifestation of positive CO anomalies in winter, as well as in early spring and late autumn, i.e. outside the fire season (see Fig. 2c). This fact can be associated with anthropogenic influence. The longer lifetime of CO in the atmosphere (from several weeks to several months) increases the contribution of long-range CO transport from urban areas in the formation of regional CO anomalies. A weaker relationship between CO in the Arctic atmosphere and forest fires in Siberia is also confirmed by the scatter diagram (Fig. 2f). The sensitivity of CO content in the Arctic troposphere to changes in the number of fires in Siberia, estimated with the use of linear regression to be equal to 0.4 ppmv/1000 natural fires.

*Long-range transport of biomass burning products from Siberian wildfires into the Arctic atmosphere*

For Russia as a whole, the connection between massive forest fires and atmospheric blockings was analyzed in a number of studies [27-29]. An analysis of the field of the geopotential height of the pressure 500 hPa level (H500) showed that episodes of extreme fire activity in Siberian forests (Fig. 2a) were associated with large-scale circulation typical of atmospheric blocking. Atmospheric blockings contribute to the formation of regional anomalies of atmospheric impurities, including CO, aerosol, and ozone (with the formation of mini-ozone holes) [27-33].

Figure 3 presents the results of the analysis of long-range BC transport to the Arctic atmosphere during strong forest fires in Siberia in the first decade of August 2019. This period was characterized by the formation of the strongest positive BC anomaly over the Russian sector of the Arctic (see Fig. 2a and 2b).

Analysis of the longitude-time variations of the blocking index (similar to the index in [34]) indicates the manifestation of atmospheric blocking in the longitude sector 95-125°E in the first decade of August 2019 (Fig. 3b). The spatial distribution of H500 (Fig. 3a) during this period was characterized by a quasi-stationary high-pressure anomaly over the central part of Siberia and two stable low-pressure anomalies over the Northern Urals and the Far East. The noted features in the H500 field are characteristic of large-scale atmospheric circulation during atmospheric blocking. An analysis of fire activity indicates that atmospheric blocking in the first decade of August 2019 was accompanied by the development of massive fires in Siberian forests (Fig. 3e). A joint analysis of forest fires and H500 anomalies shows that the main sources of atmospheric pollution by biomass combustion products are associated with the blocking area (Fig. 3a).

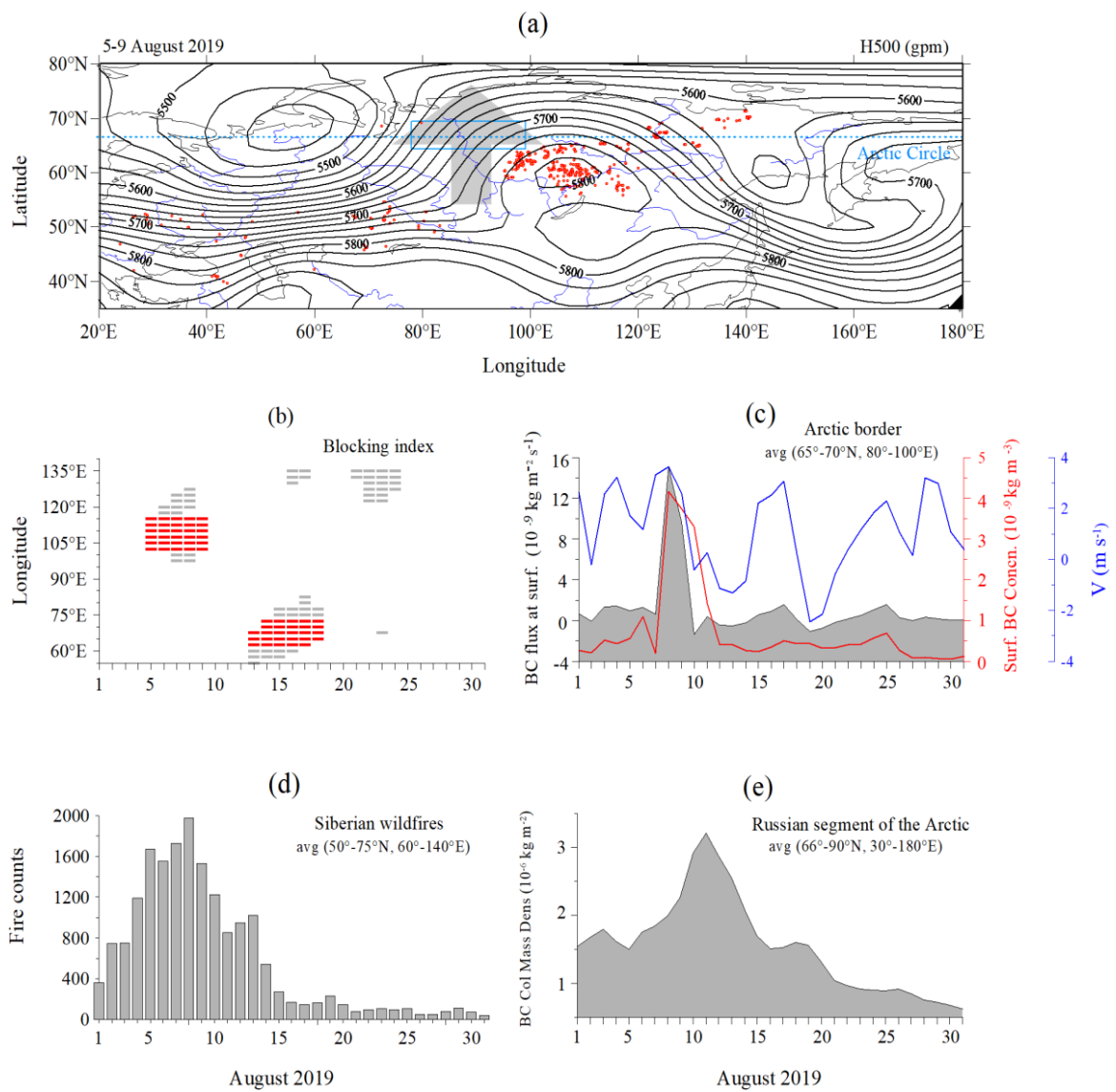

**Figure 3.** (a) Spatial distributions of H500 (black isolines) and wildfires (red circles) during 5-9 August 2019. Blue rectangle denotes territory considered in (c), thick arrow shows the northward flux of burning products. (b) Blocking index (red symbols stand for blocking conditions lasting at least 5 days). (c) Meridional BC flux in the atmospheric surface layer through the Arctic Circle (shaded); BC mass concentration (red line) and meridional wind at the 10 m height (blue line) in the atmospheric surface layer close to the Arctic Circle (see also (a)). (d) Number of Siberian wildfires. (e) BC total column over the Russian segment of the Arctic.

The long-range transport of pyrogenic air pollutants in August 2019 was associated with atmospheric blocking. Figure 3a shows that to the west of the blocking anticyclone, the western zonal flow deviated far to the north, and to the east of the blocking anticyclone, to the south, carrying air polluted by combustion products. The H500 spatial gradient to the west of the blocking anticyclone, the H500 gradient was greater than to the east of the anticyclone (Fig. 3a), and the wind speed over the western periphery of the anticyclone was higher than the wind speed over its eastern periphery. As a whole, in the area of atmospheric blocking, the flow of combustion products from south to north dominated, with the transport of combustion products from forest fires to the Arctic.

Flux of BC in the atmosphere from forest fires in Siberia to the Arctic atmosphere was estimated using

$$F_{BC} = C_{BC} \cdot V,$$

where $F_{BC}$ is the BC flux through the Arctic Circle (AC) [kg m$^{-2}$ s$^{-1}$], $C_{BC}$ is the BC mass concentration at the surface [kg m$^{-3}$], $V$ is the meridional wind at the height of 10 m [m s$^{-1}$]. Figure 3c shows day-to-day variations in the meridional AC flux, regional AC mass concentration, and meridional wind in the surface layer of the atmosphere in August 2019. The data were averaged for the region (66-69°N, 80-100°E) in the vicinity of the AC (see also Fig. 3a). In early August, during relatively weak forest fires, even despite the strengthening of the south wind, a weak flow of aircraft to the Arctic was observed near the AU. At the end of the first ten days of August 2019, forest fires in Siberia reached their maximum development, which led to the saturation of the subarctic atmosphere with combustion products (Fig. 3c). The strengthening of the south wind over the southern Arctic latitudes led to a sharp increase in the flow of aircraft to the Arctic, reaching a maximum value of 1.5 10-2 kg km-2 on August 8, 2019. With the beginning of the second ten days of August, a decrease in the number of fires and a change in wind from south to north led to a weakening and even a change in the direction of the BC flow, contributing to a decrease in the regional concentration of BC in the Arctic atmosphere (Fig. 3e). The subsequent strengthening of the south wind only led to a relatively weak increase in the flow of aircraft to the Arctic due to a significant decrease in the number of fires in Siberian forests (Fig. 3d).

Changes in the atmospheric pollution over the Russian sector of the Arctic (Fig. 3e) indicate an increase in the regional atmospheric BC abundance in the first decade of August

2019, which reached a maximum of $3.2 \cdot 10^{-2}$ kg km$^{-2}$ on August 11, 2019. There is a three-day delay between the maximum BC content in the Arctic atmosphere and the maximum flux of BC over AC, which is explained by the time required for the transport of BC from forest fires in Siberia to the Arctic.

**Conclusions**

The analysis of the causes and mechanisms of the formation of large-scale anomalies in BC and CO contents in the atmosphere of the Russian sector of the Arctic revealed a relationship of the regional BC and CO anomalies with forest fires in Siberia. The estimates of the sensitivity of regional content of BC and CO in the atmosphere of the Russian Arctic to the corresponding changes in the number of Siberian fires were obtained about $2 \cdot 10^{-2}$ kg km$^{-2}$ / 1000 fires and 0.4 ppmv / 1000 fires, respectively.

According to the analysis of fire activity in the forests of Siberia in the past two decades, there has been a predominant increase in the number of intense forest fires with a statistically significant linear trend of 1.8 ($\pm$ 0.9) %/10 years (Fig. 1a). Analysis of changes in the number of weak forest fires did not reveal any trend.

An analysis of extreme BC pollution of the high-latitude atmosphere during catastrophic forest fires in Siberia in the summer of 2019 reveals that regional features of large-scale atmospheric circulation associated with atmospheric blocking event contributed to the long-range transport of BC from the fires of the Siberian forests to the Arctic. Estimates of BC flux from Siberian forest fires through the Arctic circle into the Arctic have been obtained with the maximum BC flux in August 2019 reached the value of $1.5 \cdot 10^{-2}$ kg km$^{-2}$ s$^{-1}$.

**Acknowledgements**

This work was carried out with support of the Russian Science Foundation (project 19-17-00240).